\iffalse Before sending further, pass this check list:
  [ ] find and resolve all "??" and "\ifnote"
  [ ] Spell-check
  [ ] order of using/defining abbreviations
  [ ] citation order (use RefTest)
  [ ] figure order & reference order
  [ ] Check LaTeX  output files (*.log) for warnings
  [ ] Check BibTeX output (screen) for warnings
  [ ] update PACS
  [ ] decide on color figures
  [ ] Re-read paper in the morning

After completing this list, if you made at least one correction,
re-do this check-list from the beginning, until no corrections will
be done.\fi

\documentclass[amsmath,amssymb,prb,preprintnumbers,superscriptaddress,twocolumn]{revtex4}

\usepackage{graphicx}
\usepackage{dcolumn}    
\usepackage{bm}         
\usepackage{color}      
\usepackage{ulem}       

\newcommand{\rw}[1]{\textcolor{red}{\textsf{#1}}}

\newif\ifgraph

\graphtrue                   %

\begin{document}


\title{Improved tunneling magnetoresistance at low temperature in manganite junctions
grown by molecular beam epitaxy}

\author{R.~Werner}
\affiliation{  Physikalisches Institut -- Experimentalphysik II,  Universit\"{a}t T\"{u}bingen,   Auf der Morgenstelle 14,   72076 T\"{u}bingen, Germany }
\author{A.~Yu.~Petrov}
\affiliation{ CNR-IOM TASC National Laboratory,  S.S. 14 Km 163.5 in AREA Science Park,  34012 Basovizza, Trieste, Italy}
\author{L.~Alvarez~Mi\~{n}o}
\affiliation{Universidad Nacional de Colombia, Sede Manizales, Cra 27 \# 64-60 Manizales, Colombia}
\author{R.~Kleiner}
\author{D.~Koelle}
\affiliation{  Physikalisches Institut -- Experimentalphysik II,  Universit\"{a}t T\"{u}bingen,   Auf der Morgenstelle 14,   72076 T\"{u}bingen, Germany }
\author{B.~A.~Davidson}
\email{davidson@tasc.infm.it}
\affiliation{ CNR-IOM TASC National Laboratory,  S.S. 14 Km 163.5 in AREA Science Park,  34012 Basovizza, Trieste, Italy}
\date{\today}

\begin{abstract}
We report resistance versus magnetic field measurements for a
La$_{0.65}$Sr$_{0.35}$MnO$_3$/SrTiO$_3$/La$_{0.65}$Sr$_{0.35}$MnO$_3$ tunnel junction grown by molecular-beam epitaxy, that show a large field window of extremely high tunneling magnetoresistance (TMR) at low temperature. 
Scanning the in-plane applied field orientation through $360^{\circ}$, the TMR shows 4-fold symmetry, i.e. biaxial anisotropy, aligned with the crystalline axes but not the junction
geometrical long axis. 
The TMR reaches $\sim 1900\,\%$ at 4\,K, corresponding to an interfacial spin
polarization of $>95\,\%$ assuming identical interfaces. 
These results show that uniaxial anisotropy is not necessary for large TMR, and lay the groundwork for future improvements in TMR in manganite junctions.
\end{abstract}
%
%
\maketitle
%

The figure of merit for magnetic tunnel junctions (MTJs) is the tunneling magnetoresistance (TMR) ratio, which determines its performance in practical devices such as magnetic random access memories (MRAMs) and low-field sensors\cite{Wolf01}.
An MTJ consists of two ferromagnetic electrodes separated by a thin insulating tunneling barrier.
According to the Julliere model \cite{Julliere75}, the TMR ratio is defined as ${\rm TMR_J}=(R_{\mathrm{AP}} - R_P)/R_P = 2P_1P_2/(1-P_1P_2)  $.
Here $P_1$ and $P_2$ are the spin polarizations of the two electrodes and $R_{\mathrm{AP}}$ and $R_P$ are the junction resistances with antiparallel and parallel orientation of magnetization $M$ of the two electrodes, respectively.
Accordingly, an MTJ made from half-metallic electrode materials, such as doped manganites \cite{Coey99}, should yield an infinite TMR ratio at temperature $T$ well below the Curie temperature $T_C$.
Noting that TMR is more precisely associated with the properties of the electrode/barrier interface \cite{Teresa99}, this concept has been extended to also describe interfaces as half-metallic \cite{Bowen07}, i.e.~the TMR is determined by the spin-polarization of the local density of states at the two interfaces with the barrier.
Ferromagnetic correlations at manganite surfaces and interfaces are known to be weaker than in bulk, causing a ''dead layer`` \cite{Sun99, Bibes01, Tebano08}.
For example, at the vacuum/La$_{1-x}$Sr$_x$MnO$_3$ (LSMO) interface the nonferromagnetic layer is about 3 unit cells (uc) thick at $T=200\,$K \cite{Verna10}, well below the bulk $T_C \approx 360$\,K for ferromagnetic LSMO with optimal doping $x = 0.35$ (F-LSMO).
This and other effects have been discussed  to explain the disappearance of the TMR well below the bulk $T_C$ in manganite MTJs \cite{Sun99, Viret97, Jo00, Donnell00}.
Attempts have been made to ''engineer`` the interfaces  by creating a doping profile  to overcome this problem, and even though the TMR ratio remained low\cite{Yamada04}, spectroscopic characterization  suggested  this approach could improve the low-temperature TMR\cite{Kavich07}.

To date, TMR at small dc voltage bias of MTJs based on nonoxide electrodes reached $ \approx 1150\,\%$ at $T=5\,$K \cite{Ikeda08} while the highest ratio was reported for manganite/titanate interfaces with a maximum value of about $1800\,\%$ at $4\,$K in a very small window of applied in-plane magnetic field $H$ \cite{Bowen03}.
Here, an antiferromagnetic CoO layer was used to pin the upper electrode via exchange bias,\cite{Schuller04} that can favor uniaxial anisotropy in the pinned electrode; such anisotropy was claimed  necessary for the stabilization of well-defined antiparallel states and high TMR ratios\cite{Bowen03, Jo00}.

In this letter, we report on the TMR of MTJs based on F-LSMO, with an antiferromagnetic $x=0.65$ LSMO exchange bias layer (AF-LSMO) and a SrTiO$_3$ (STO) barrier, grown by molecular beam epitaxy.
We find a TMR ratio up to $\sim 1900\,\%$ at $T=4\,$K, which  decreases rapidly with increasing $T$, disappearing  at $\sim 280\,$K.
Rotating the applied in-plane magnetic field, we find a four-fold symmetry of the TMR, indicating that uniaxial anisotropy in not required for high TMR ratios.

For sample fabrication, we developed atomic-layer control of LSMO and STO growth and their interfaces by combining reactive molecular beam epitaxy (MBE) \cite{Eckstein95} with \textit{in situ} reflection high-energy electron diffraction (RHEED) techniques, extending the work of Haeni {\it et al.} \cite{Haeni00}.
These RHEED techniques permit us to adjust the surface termination at any point during deposition, including during interface growth \cite{Davidson_unpub}.
LSMO/STO/LSMO trilayers were grown on (001)-oriented STO substrates at a typical substrate temperature $T_s = 750^{\circ}$\,C and ozone pressure $p = 10^{-6}$\,mbar.
The bottom and top F-LSMO electrode thicknesses were $50\,$uc, separated by a tunnel barrier of stoichiometric SrTiO$_3$, $5-6\,$uc thick.
A 100 uc thick AF-LSMO layer was grown underneath the bottom electrode to increase and shift  its coercive field $H_c$ due to exchange bias.
The resulting difference in $H_c$ between the electrodes favors the establishment of fully antiparallel magnetization orientation of the two electrodes in a larger window of $H$ \cite{Schuller04} in the resistance versus magnetic field $R(H)$ loops used to determine the TMR ratio.
The effect of exchange bias on the junction TMR characteristics should only be seen below $\sim 250\,$K, in agreement with exchange bias effects seen in the hysteresis loops of an AF-LSMO (100 uc)/F-LSMO (50 uc) bilayer by SQUID magnetometry measured independently. 

%
\begin{figure}
  \includegraphics[width=0.48\textwidth]{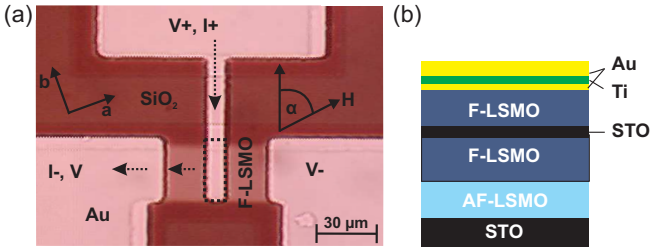}
  \caption{(Color online) \rw{(a)} Optical image of the MTJ. The arrows indicate the current flow, while the junction area is indicated by the dashed rectangle. \rw{(b)} The inset shows a cross section of the stacking sequence of the MTJ.}
  \label{fig1}
\end{figure}


Vertical mesa MTJs were patterned in several steps by photolithography and Ar ion milling; for details see Ref.~[\onlinecite{Patterning}].
Figure \ref{fig1} shows in (a) an optical microscope image of an MTJ with a $5 \times 30\,\mu{\rm m}^2$ mesa with vertical current injection from a Au/Ti/Au top contact and the stacking sequence of the sample in (b).
The junction resistance was more than an order of magnitude larger than the electrode resistances, ensuring uniform injection of bias current $I$~ \cite{Veerdonk97}.
Electrical transport measurements were made in 2- and 3-point geometries in a He-flow cryostat at $T=4-300$\,K.
A Helmholtz coil outside the cryostat allows full rotation of $H$ from $\alpha = 0-360^{\circ}$ with a maximum amplitude $ \mathrm{H_{\mathrm{max}}} = 1$\,kOe.
The angle $\alpha$ describes the relative orientation of $H$ with respect to the long side of the junction (c.f.~Fig.~\ref{fig1}).
All $R(H)$ loops shown or discussed below were taken after the junction was field-cooled at $ H=H_{max}$ with fixed cooling angle $\alpha_{FC}=70^\circ$, which corresponds to the crystallographic lattice $a$-axis direction.
The (differential) junction resistance was measured by a lock-in amplifier at a dc voltage bias $V=0$, using an ac current amplitude of a few nA.
Full characterization of the junction $I$($V$) and TMR as a function of dc voltage bias will be reported separately.

\begin{figure}[h]
  \includegraphics[width=0.48\textwidth]{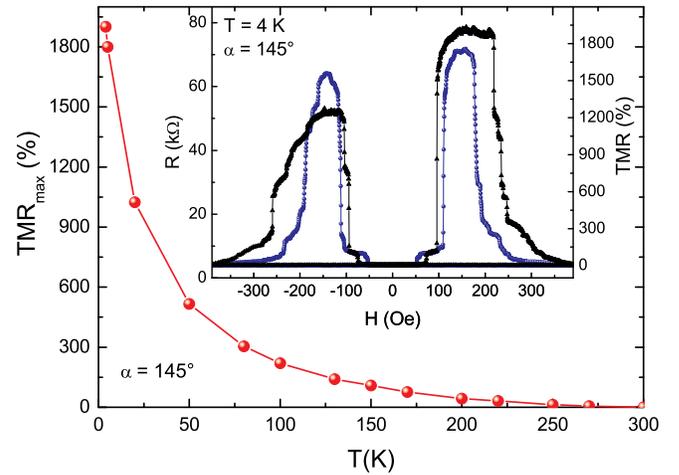}
  \caption{Maxmimum TMR ratio vs. temperature for $\alpha_{FC} = 70^{\circ}$, $\alpha = 145^{\circ}$ .
  The inset shows two $R(H)$ sweeps measured at $ 4$\,K  for the same field conditions  after  different  cooling cycles .}
  \label{fig2}
\end{figure}

The inset of Fig.~\ref{fig2} shows two representative $R(H)$ loops at low temperature ($T=4$\,K) and $\alpha = 145^\circ$, taken after two identical cooling cycles.
We define ${\rm TMR}(H)=(R(H)-R_\mathrm{min})/R_\mathrm{min}$ and the maximum TMR ratio within an $R(H)$ loop as ${\rm TMR_{\rm max}}=(R_\mathrm{max}-R_\mathrm{min})/R_\mathrm{min}$. Here, $R_\mathrm{max}$ and $R_\mathrm{min}$ are the maximum and minimum junction resistances, and only in the ideal case of uniform and fully antiparallel electrode magnetization does this definition coincide with TMR$_J$ defined above.
The $R(H)$ loops differ in shape (asymmetry), $R_\mathrm{max}$ and width $\Delta H$ of the magnetic field window, defined as the FWHM of the $R(H)$ peaks.
The $R(H)$ loop with larger $\Delta H$ shows a stronger asymmetry in peak heights and the highest $\rm TMR_{\rm max} = 1904\,\%$, which also exceeds measured TMR ratios for other combinations of field orientations $\alpha_\mathrm{FC}$ and $\alpha$, during cooling and measurement, respectively.

The $R(H)$ loops (c.f. inset of Fig.~\ref{fig2}) show switching to the high and low R states at $\sim 100\,$Oe and $200-300\,$Oe, which we attribute to $H_c$ of the upper (free) and bottom (exchange-biased) electrodes, respectively.
We note that these values of increased H$_c$  as compared to single layer F films   and nearly negligible hysteresis loop shifts $H_{eb}$ for the bottom electrode due to an exchange bias field deduced from the inset of Fig.~\ref{fig2}, are consistent with $H_{eb}$ and $H_c$   in exchange bias AF/F bilayers of the same thicknesses measured separately (not shown).
The $R(H)$ loops for nominally identical cooling cycles implies that the exchange bias coupling at the AF/F interface can vary between cooling cycles.
We attribute this to a history dependence of the exchange bias, i.e., the exchange coupling at the AF/F interface may depend sensitively on exactly its cooling history.
Reproducible exchange coupling is required for practical devices, and should be further investigated in our LSMO heterostructures.

The main panel in Fig.~\ref{fig2} shows the temperature dependence of $\rm{TMR}_\mathrm{max}$ which decays quickly with increasing $T$ and vanishes at  $\sim 280\,$K, which is also near  temperature at which exchange bias effects disappear.
The strong decay of $\mathrm{TMR}_\mathrm{max}(T)$, could in principle be explained by different $T-\,$dependent mechanisms, for example intrinsic to the LSMO/STO interfaces (such as intrinsic loss of spin polarization) or extrinsic (such as weakening of the exchange bias pinning of the bottom electrode).
Any study of the temperature dependence of the domain structure and thus relative magnetization orientations of the electrodes would require a microscopic technique such as demonstrated in Ref. [\onlinecite{Wagenknecht06}].
Evidence for weakened exchange bias pinning at higher $T$ is seen in the decreasing asymmetry of $\Delta H$ and $R_{max}$ as temperature is increased: the asymmetry in $R(H)$ between positive and negative $H$ disappears around $\sim 100\,K$.
Further characterization is necessary to distinguish between these competing mechanisms, and will be crucial to understand any limits to the potential high-temperature TMR.
%
\begin{figure}[h]
  \includegraphics[width=0.48\textwidth]{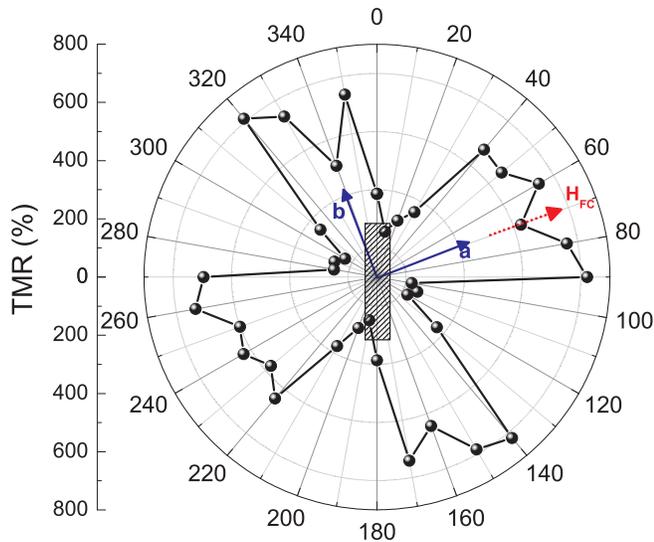}
  \caption{Polar plot of the maximum TMR ratio (at $T=30\,$K) vs.~field direction $\alpha$ for field-cooling at $\alpha_{FC} = 70^{\circ}$ (red dashed arrow). Crystalline axis are indicated by blue arrows; junction orientation is indicated by rectangle.}
  \label{fig3}
\end{figure}

Figure \ref{fig3} shows a polar plot of TMR$_\mathrm{max}(\alpha)$ at $T=30\,$K, after field-cooling along the $a$-axis of LSMO, which still shows a slight asymmetry for opposite field directions.
The 4-fold symmetry of TMR$_\mathrm{max}(\alpha)$ indicates a biaxial anisotropy, with easy axis along the $a$- and $b$ directions of our F-LSMO layers.
The slight difference in TMR$_\mathrm{max}$ values for orientations close to the $a$- and $b$-axes could be due to, e.g., a shape effect correlated to the junction long  axis, or a small anisotropy in the exchange bias ; 4-fold symmetry in the switching fields has been previously reported \cite{Jo02}.
We note that cooling the device with the field oriented in different directions does not change the 4-fold symmetry seen in the polar plot, although it can have a sizable impact on TMR$_\mathrm{max}$.
Additional study of the interplay between junction geometry and magnetocrystalline anisotropy will be necessary to further optimize these MTJs.


In summary, we have shown $R(H)$ at $4\,$K for a manganite MTJ with a useful  TMR$_{\mathrm{max}}$ ratio of 1900\%, the largest value for any MTJ reported so far  in the literature at low dc bias.
The strength of pinning of one electrode magnetization via exchange bias has a noticeable influence on the TMR both by enlarging the magnetic field window of antiparallel alignments and inducing an asymmetry in the $R(H)$ curves, but is also very sensitive to the cooling history.
The polar plot of  TMR$_{\mathrm{max}}(\alpha)$ demonstrates that uniaxial anisotropy in the F layers is not necessary for high TMR.
It is reasonable that interface roughness, oxygen vacancies and the interface growth play a crucial role in the exchange bias mechanism in these manganite interfaces as has been demonstrated in more conventional EB systems \cite{Kuch06}, and merits further study.
%

%
R.~Werner gratefully acknowledges support by the Cusanuswerk, Bisch\"{o}fliche Studienf\"{o}rderung.
B. A. D. and A. Yu. P. acknowledges support by the FVG Regional project SPINOX funded by Legge Regionale 26/2005 and Decreto 2007/LAVFOR/1461.
L. Alvarez-Mi\~{n}o thanks the Abdus Salam ICTP (Trieste) for financial support through a STEP fellowship.
The TASC Technical group is acknowledged for contributions to the design and construction of the MBE system. 
This work was funded in part by the Deutsche Forschungsgemeinschaft (Project KO 1303/8-1).

\clearpage

\section*{\Large Supplementary information }

This supplement describes in detail the patterning procedure for MTJ fabrication, which is based on optical lithography and Ar ion milling.
We use a self-alignment process for final definition of the MTJ area, which ensures alignment of the metallic wiring layer with the upper electrode of the MTJ and which prevents exposure of the junction edges to chemicals during any time in the patterning process.
The MTJs were patterned in four steps, which we denote as (I) feed line patterning, (II) edge isolation, (III) metallization and (IV) junction milling.

During  Ar$^+$-milling (beam energy $0.3\,$keV), the sample was mounted on a water-cooled ($T \approx 8^\circ$C) copper block which reduces heating of the sample.
A shutter in front of the sample was used to mill in intervals of $t_o=5\,$s open and $t_c=10\,$s closed.
This helps to avoid excessive heating of the sample surface, which might induce interdiffusion at the interface and oxygen loss in the LSMO and STO layers, both of which do have a detrimental effect on the TMR.

\vspace*{-5mm}
\section[I]{Feed line patterning}

\begin{figure}[b]
\includegraphics[width=0.9\columnwidth]{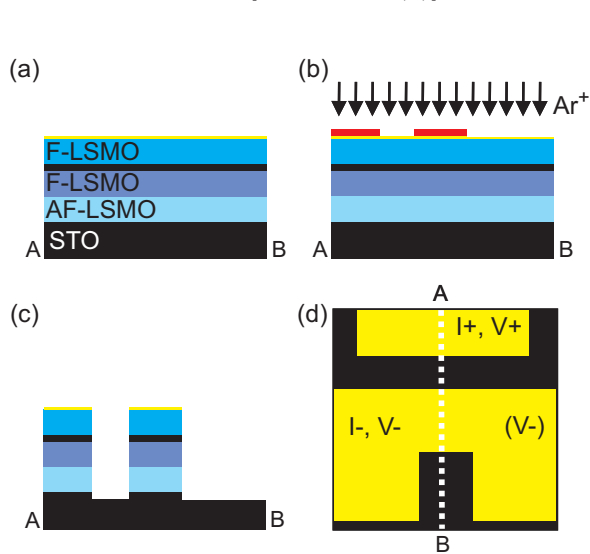}
\caption{\textbf{Feed line patterning:}
cross sections of the sample, (a) prior to patterning, (b) before, and (c) after Ar ion milling and removal of PR. (d) shows top view corresponding to (c); white dashed line indicates the position of cross sections $\overline{AB}$ shown in (a)-(c).}
\label{figS1}
\end{figure}

This step defines the feed lines of the MTJ.
Figure~\ref{figS1}~(a) shows a cross section of the as-grown sample, indicating the stacking sequence.
The two F-LSMO electrodes are separated by 6 uc thick STO barrier layer (black), and the sample was covered in-situ by a $\sim 1\,$nm thick Au layer (yellow), for protection against humidity and chemicals during the patterning process.
In a first step, the sample was coated by photoresist (PR) which was patterned by optical lithography [c.f. Fig.~\ref{figS1}~(b)].
Subsequently, the sample was loaded into the milling chamber, and the regions exposed to the Ar ions were milled down into the STO substrate.
Figure~\ref{figS1}~(c) shows the cross section after PR removal in acetone, and Fig.~\ref{figS1}~(d) shows the corresponding top view.
After step I, the feed lines for current and voltage terminals ($I^+$, $V^+$) and ($I^-$, $V^-$) are completely separated.

\vspace*{-5mm}
\section[II]{Edge isolation}

\begin{figure}[b]
\includegraphics[width=0.9\columnwidth]{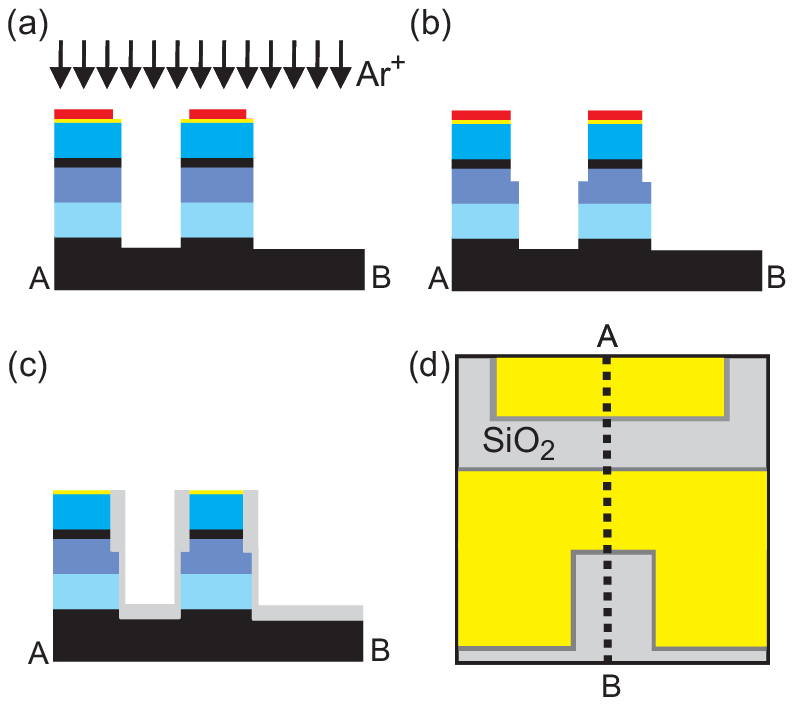}
\caption{\textbf{Edge isolation:}
cross sections of the sample, (a) before, and (b) after Ar ion milling into the bottom F-LSMO layer. (c) shows the structure after deposition of an SiO$_2$ layer and lift-off patterning (removal of PR). (d) shows top view corresponding to (c); black dashed line indicates the position of cross sections $\overline{AB}$ shown in (a)-(c).}
\label{figS2}
\end{figure}

The next step provides isolation of the edges of the feed lines by an SiO$_2$ layer and at the same time defines the length (along $\overline{AB}$) of the MTJ.
Here, a mask was used which had essentially the same geometry as the one used for feed line patterning.
However, the structures on this mask are slightly smaller, i.e.~their edges are shifted by $10\,\mu$m with respect to the previous mask; c.f.~Fig.~\ref{figS2}(a).
Furthermore, Ar ion milling is stopped after milling into the bottom F-LSMO electrode as indicated in Fig.~\ref{figS2}~(b).
Subsequently, we sputtered $\sim 40\,$nm of SiO$_2$ in a Ar/O$_2$ atmosphere (at 200\,Pa), followed by lift-off.
This process results in the structure shown in Figs.~\ref{figS2}~(c) and (d).

\section[III]{Metallization}

Step III provides a low-resistance connection between the upper F-LSMO electrode of the MTJ and the feed line with the ($I^+$, $V^+$) terminals, which shall ensure homogeneous current injection into the MTJ.
For this purpose, a Ti(2\,nm)/Au(50\,nm)/Ti(10\,nm) trilayer is deposited in-situ by electron beam evaporation.
The bottom Ti layer is used as an adhesive layer underneath the Au layer, in order to improve sticking on the SiO$_2$ layer.
The upper Ti layer on top of the Au layer will be used as a milling mask in step IV.
We note that the in-situ Au layer (on top of F-LSMO) is crucial to avoid a high contact resistance between the upper F-LSMO layer and the bottom Ti layer.

\begin{figure}[b]
\includegraphics[width=0.9\columnwidth]{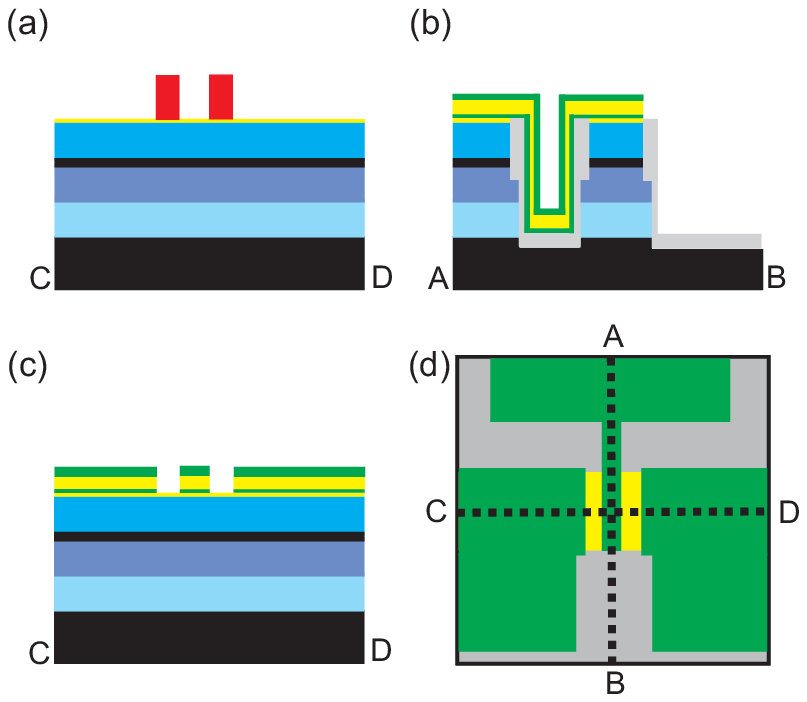}
\caption{\textbf{Metallization:}
cross sections of the sample, (a) after patterning a lift-off mask, and (b),(c) after Ti/Au/Ti trilayer deposition and lift-off patterning (removal of PR). (d) shows top view corresponding to (b) and (c); black dashed lines indicate the position of the cross sections $\overline{AB}$ and $\overline{DC}$ shown in (a), (b) and (c).}
\label{figS3}
\end{figure}

Step III starts with preparation of a lift-off mask [c.f.~Fig.~\ref{figS3}~(a)], which covers everywhere, except the green areas shown in Fig.~\ref{figS3}~(d).
Subsequently, the Ti/Au/Ti trilayer is deposited and patterned by lift-off.
Figures \ref{figS3}~(b) and (c) show vertical ($\overline{AB}$) and horizontal ($\overline{CD}$) cross sections, respectively, after this step.
Figure~\ref{figS3}~(d) shows the corresponding top view at the end of this metallization process.
The width of the central Ti/Au/Ti strip (along $\overline{CD}$) defines the width of the MTJ.

\vfill

\section[IV]{Junction milling}

\begin{figure}[h]
\includegraphics[width=0.9\columnwidth]{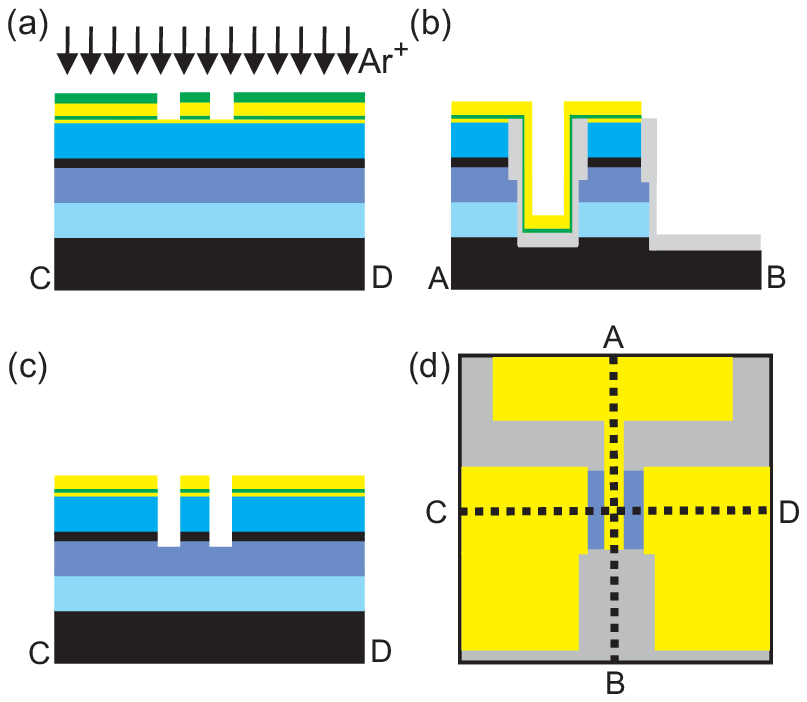}
\caption{\textbf{Junction milling:}
cross sections of the sample, (a) at the start of Ar milling, and (b),(c) after Ar milling. (d) shows top view corresponding to (b) and (c); black dashed lines indicate the position of the cross sections $\overline{AB}$ and $\overline{DC}$ shown in (a), (b) and (c).}
\label{figS4}
\end{figure}

Step IV is the final patterning step of the MTJ, using a self-alignment method.
Here, the whole sample area is milled with Ar$^+$, without any PR [c.f Fig.~\ref{figS4}~(a)].
This step is possible, as the milling rate for Ti is much lower than the ones for Au, F/AF-LSMO and STO.
During milling, the areas uncovered by Ti are milled down to the bottom F-LSMO electrode as shown in Fig.~\ref{figS4}~(b).
The thickness $\sim 10\,$nm of the Ti layer has been adjusted to ensure that after milling, also the upper Ti layer has been removed completely [c.f.~Figs.~\ref{figS4}~(b) and (c)].
The final top view, shown in Fig.~\ref{figS4}~(d), is a schematic representation of the optical image shown in Fig.~1 in the main paper.
%
%
\clearpage
\end{document}